# Solutions of the 2D Schrodinger equation and its thermal properties for improved Ultra Generalized Exponential Hyperbolic potential (IUGE-HP)

Akpan Ndem Ikot[1*], Collins Okon Edet[1], Uduakobong Sunday Okorie[1&3], Abdel-Haleem Abdel-Aty[4&5], M. Ramantswana[2], Gaotsiwe Joel Rampho[2], Nawal A. Alshehri[6], S.K Elagan[6] and Savas Kaya[7]

[1]Theoretical Physics Group, Department of Physics, University of Port Harcourt, P M B 5323 Choba , Port Harcourt –Nigeria

[2]Department of Physics, University of South Africa, Florida 1710, Johannesburg-South Africa

[3]Department of Physics, Akwa Ibom Stae University ,Ikot Akpaden, Uyo-Nigeria

[4]Department of Physics, Colleage of Sciences, University of Bisha,P.O.Box 344,Bisha 61922,Saudi Arabia

[5]Physics Department, Faculty of Science, Al-Azhar University, Assiut,71524, Egypt

[6]Department of Mathematics and Statistics, College of Science, Taif University, P.O.Box 11099, Taif 21944, Saudi Arabia

[7] Science Faculty, Department of Chemistry,Cumhuriyet University,Sivas,58140,Turkey

*email: ndemikotphysics@gmail.com

**ABSTRACT**

In this research work, we investigate the 2D Schrodinger equation with improved ultra-generalized exponential-hyperbolic potential (IUGE-HP) is scrutinized taking into consideration the effects external magnetic and Aharanov-Bohm (AB) fields within the non-relativistic quantum mechanics regime. The Schrodinger equation in 2D for this consideration is solved using the asymptotic iteration method (AIM) and the energy equation and eigenfunction for the IUGE-HP in the presence of magnetic and AB fields are obtained in a compact form. The numerical energy spectra is analyzed and it is observed that the presence of the external fields eliminates degeneracy from the spectra and the different values of the control parameter causes shifts in the energy spectra. The thermal and magnetic properties of the IUGE-HP in the presence of this external fields are duly investigated. The variations of the thermal and magnetic properties with control parameter are analyzed for negative $(\tau < 0)$ and positive $(\tau > 0)$. The present research finds applications in condensed matter, molecular and atomic physics respectively.

1. **Introduction**

In non-relativistic quantum mechanics regime, the Schrödinger equation (SE) is known to provide the basic information that aids in understanding the behaviour of a given quantum system [1-4]. Different researchers have devoted much attention to the studies of the SE involving various potentials [5–13]. We propose a potential named the improved ultra-generalized exponential-hyperbolic potential (IUGE-HP) in furtherance of the understanding of SE. This potential is defined as:

$$V(r) = \frac{Ae^{-4(\alpha+\delta)r} + Be^{-2(\alpha+\delta)r}}{r^2} + \frac{Ce^{-2(\alpha+\delta)r} - D\left[e^{-(\alpha+\delta)r}\cosh(\alpha+\delta)r + \frac{\tau}{2}\right] + G\cos ech(\alpha+\delta)re^{(\alpha+\delta)r}}{r} + K \quad (1).$$

Here, $A, B, C, D, G$ and $K$ represent different potential parameters; $\alpha$ and $\delta$ are screening parameters. These screening parameter may be either the same or different, depending on the system considered; $\tau$ is known as the control parameter that can either be negative $(\tau < 0)$, positive $(\tau > 0)$ and zero. The IUGE-HP of Eq. (1) is a generalization of the ultra-generalized exponential-hyperbolic potential [14]. For instance, if $A = B = C = G = 0, \alpha = \delta = 0, \tau = 1$ then the IUGE-HP turns to Coulomb potential $V(r) = -\frac{D}{r}$. Similarly, if $A = B = C = G = 0, \alpha = \delta, \tau = 0$, then the IUGE-HP turns to screened Coulomb potential, $V(r) = -D\frac{e^{-2\alpha r}}{r}$ among others.

In recent developments, many research articles considering the magnetic and Aharanov-Bohm (AB) fields effects on several quantum mechanical systems have been published. For instance, Ikot et al. [15] solved the SE involving both the screened Kratzer potential (SKP) and external magnetic and AB fields using the factorization method. The authors evaluated magnetization and magnetic susceptibility of the system at zero and finite temperatures. Also, different thermodynamic functions of the system considered were discussed. By employing the superstatistics formalism, Ikot and his collaborators [16] studied the effects of magnetic and AB fields on the thermodynamic functions of pseudo-harmonic potential for selected diatomic molecules. Edet et al. [17] used the NUFA method to obtain solutions of the SE involving both Hellmann potential and magnetic and AB fields. It was observed that degeneracies can be eliminated better with the use of AB field, as compared to the magnetic field. Rampho et al. [18] studied the SE involving improved screened Kratzer potential (ISKP) with magnetic and Aharanov-Bohm (AB) fields. Others studies carried out by several authors can be seen in Ref. [19-34].

In the present work, we solve the SE with the IUGE-HP, together with the magnetic and AB flux fields via the asymptotic iteration method (AIM). The energy spectra of IUGE-HP obtained will be used to calculate the partition function and other thermodynamic properties such as entropy, mean free energy, specific heat capacity and magnetic susceptibility. The effects of the control parameter on the energy spectra, the thermal and magnetic properties of the IUGE-HP will be discussed. The organization of this article is given as follows. In section 2, we solve of the 2D SE with the improved ultra-generalized exponential-hyperbolic potential (IUGE-HP) under the combined effects of external magnetic and AB flux fields. In section 3, the effects of the control parameter on the thermodynamics properties of IUGE-HP in the presence of external fields are discussed. Results obtained are presented and discussed in section 4. Finally, we present a the concluding remarks of the work in section 5.

2. **Schrödinger Equation of the IUGE-HP with Magnetic and AB Fields**

The Hamiltonian operator of a charged particle, subjected to move in the IUGE-HP by the help of both magnetic and AB fields can be first represented in cylindrical coordinates. Thus, the SE for this consideration is written as follows [16, 18, 34];

$$\left[\frac{1}{2\mu}\left(i\hbar\vec{\nabla} - \frac{e}{c}\vec{A}\right)^2 + \frac{Ae^{-4(\alpha+\delta)r} + Be^{-2(\alpha+\delta)r}}{r^2} + \frac{Ce^{-2(\alpha+\delta)r} - D\left[e^{-(\alpha+\delta)r}\cosh(\alpha+\delta)r + \frac{\tau}{2}\right] + G\cos ech(\alpha+\delta)re^{(\alpha+\delta)r}}{r} + K\right]\psi(r,\varphi) = E_{nm}\psi(r,\varphi) \quad (2),$$

where $E_{nm}$ denotes the energy of the system; $\mu$ being the reduced mass of the system; the vector "$\vec{A}$" being represented as a superposition of two terms $\vec{A} = \vec{A}_1 + \vec{A}_2$, having the azimuthal components [16, 18] and external magnetic field with $\vec{\nabla}\times\vec{A}_1 = \vec{B}, \vec{\nabla}\times\vec{A}_2 = 0$; $\vec{B}$ is the magnetic field; $\vec{A}_1 = \frac{\vec{B}e^{-2(\alpha+\delta)r}}{(1-e^{-2(\alpha+\delta)r})}\hat{\varphi}$ and $\vec{A}_2 = \frac{\phi_{AB}}{2\pi r}\hat{\varphi}$ represent the additional magnetic flux $\phi_{AB}$ created by a solenoid with $\vec{\nabla}.\vec{A}_2 = 0$ [16, 18]. The vector potential can then be written as [16, 18];

$$\vec{A} = \left(0, \frac{\vec{B}e^{-2(\alpha+\delta)r}}{(1-e^{-2(\alpha+\delta)r})} + \frac{\phi_{AB}}{2\pi r}, 0\right) \quad (3).$$

Also, we assume a wave function in the cylindrical coordinates to be $\psi(r,\varphi) = \frac{1}{\sqrt{2\pi r}}e^{im\varphi}\rho_{nm}(r)$, where $m$ denotes the magnetic quantum number. By inserting the assumed wave function and the vector potential of Eq. (3) into Eq. (2), a radial second-order-like differential equation is obtained of the form:

$$R_{nm}''(r) + \frac{2\mu}{\hbar^2}\left[\begin{array}{l}E_{nm} - \left(\dfrac{Ae^{-4(\alpha+\delta)r} + Be^{-2(\alpha+\delta)r}}{r^2} + \dfrac{Ce^{-2(\alpha+\delta)r}}{r} - \dfrac{D/2\, e^{-2(\alpha+\delta)r}}{r} - \dfrac{D/2\,(1+\tau)}{r}\right) \\ + \dfrac{2G}{\left(1-e^{-2(\alpha+\delta)r}\right)r} + K \\ \dfrac{\hbar e\vec{B}}{\mu c}\left(m + \dfrac{\phi_{AB}}{\phi_0}\right)\dfrac{\vec{B}e^{-2(\alpha+\delta)r}}{\left(1-e^{-2(\alpha+\delta)r}\right)r} - \left(\dfrac{e^2 \vec{B}^2}{2\mu c^2}\right)\dfrac{e^{-4(\alpha+\delta)r}}{\left(1-e^{-2(\alpha+\delta)r}\right)^2} - \dfrac{\hbar^2}{2\mu}\dfrac{\eta_m}{r^2}\end{array}\right]R_{nm}(r) = 0 \quad (4),$$

where $\eta_m = (m+\xi)^2 - \dfrac{1}{4}$, $\xi = \dfrac{\phi_{AB}}{\phi_0}$ is an integer with the flux quantum $\phi_0 = \dfrac{hc}{e}$.

Eq. (4) is not exactly solvable due to the existence of a centrifugal term. Therefore, we employ the Greene and Aldrich approximation scheme [35] to overcome the centrifugal term. This approximation scheme is given;

$$\frac{1}{r^2} = \frac{4(\alpha+\delta)^2}{\left(1-e^{-2(\alpha+\delta)r}\right)^2} \qquad (5).$$

We point out here that this approximation is only valid for small values of the screening parameter $\alpha$. By considering the transformation of the form $h = e^{-2(\alpha+\delta)r}$, Eq.(4) now becomes:

$$h(1-h)R_{nm}''(h) + (1-h)R_{nm}'(h) + \frac{\begin{bmatrix}-(\varepsilon_{nm} + V_0 - V_2 + V_3 + U_1)h^2 + (2\varepsilon_{nm} - V_2 - V_1 + V_3 - V_4 - U_0)h \\ -(\varepsilon_{nm} - V_4 + V_5 + \eta_m)\end{bmatrix}R_{nm}(h)}{h(1-h)} = 0 \quad (6)$$

where

$$-\varepsilon_{nm} = \frac{\mu(E_{nm}-K)}{\hbar^2(\alpha+\delta)^2},\ V_0 = \frac{2\mu A}{\hbar^2},\ V_1 = \frac{2\mu B}{\hbar^2},\ V_2 = \frac{\mu C}{\hbar^2(\alpha+\delta)},\ V_3 = \frac{\mu D/2}{\hbar^2(\alpha+\delta)},\ V_4 = \frac{\mu D/2\,(1+\tau)}{\hbar^2(\alpha+\delta)},\ V_5 = \frac{2\mu G}{\hbar^2(\alpha+\delta)},$$

$$U_0 = \left(\frac{e\vec{B}m}{\hbar(\alpha+\delta)c} + \frac{e\vec{B}\phi_{AB}}{\hbar(\alpha+\delta)c\phi_0}\right),\ U_1 = \left(\frac{e\vec{B}}{2\hbar(\alpha+\delta)c}\right)^2 \qquad (7).$$

By employing a wave function $R_{nm}(h)$ of the form:

$$R_{nm}(h) = h^H (1-h)^J f_{nm}(h) \qquad (8),$$

where

$$J = \frac{1}{2} + \sqrt{\frac{1}{4} + V_0 + V_1 + V_5 + U_0 + U_1 + \eta_m}$$
$$H = \sqrt{\varepsilon_{nm} - V_4 + V_5 + \eta_m} \qquad (9),$$

Eq. (6) now becomes a hypergeometric differential equation of the form:

$$f''_{nm}(h) = \frac{(2H+2J+1)h-(2H+1)}{h(1-h)} f'_{nm}(h) + \frac{\left((H+J)^2 - \left(\sqrt{\varepsilon_{nm}+V_0-V_2+V_3+U_1}\right)^2\right)}{h(1-h)} f_{nm}(h) = 0 \quad (10).$$

The solution of Eq. (10) can be obtained by using the AIM [36] formalism. The methodical technique of the AIM begins now by rewriting Eq. (10) in the form [37];

$$f''_{nm}(s) - \lambda_0(h) f'_{nm}(s) - s_0(h) f_{nm}(s) = 0 \qquad (11),$$

where

$$\lambda_0(h) = \frac{(2H+2J+1)h-(2H+1)}{h(1-h)} ; s_0(h) = \frac{\left((H+J)^2 - \left(\sqrt{\varepsilon_{nm}+V_0-V_2+V_3+U_1}\right)^2\right)}{h(1-h)} \qquad (12).$$

The primes of the function $f_{nm}(h)$ in equation (34) signifies the derivatives with respect to $s$. The asymptotic feature of the method for sufficiently large $k$ is given as [36, 37]

$$\frac{s_k(h)}{\lambda_k(h)} = \frac{s_{k-1}(h)}{\lambda_{k-1}(h)} = \alpha(h) \qquad (13),$$

where

$$\lambda_k(h) = \lambda'_{k-1}(h) + s_{k-1}(h) + \lambda_0(h)\lambda_{k-1}(h),$$
$$s_k(h) = s'_{k-1}(h) + s_0(h)\lambda_{k-1}(h), \qquad (14).$$

Eq. (14) is referred to as the recurrence relations [36, 37]. In accordance with AIM [36, 37], the equation we seek can be obtained from the roots of the following equation [36]:

$$\delta_k(h) = \begin{vmatrix} \lambda_k(h) & s_k(h) \\ \lambda_{k+1}(h) & s_{k+1}(h) \end{vmatrix} = 0, \quad k=1,2,3... \qquad (15).$$

Thus, we can easily obtain the following simple arithmetic progressions:

$$\delta_0(h) = \begin{vmatrix} \lambda_0(h) & s_0(h) \\ \lambda_1(h) & s_1(h) \end{vmatrix} = 0 \Leftrightarrow H_0 = -0 - J \pm \sqrt{\varepsilon_{nm}+V_0-V_2+V_3+U_1}$$

$$\delta_1(h) = \begin{vmatrix} \lambda_1(h) & s_1(h) \\ \lambda_2(h) & s_2(h) \end{vmatrix} = 0 \Leftrightarrow H_1 = -1 - J \pm \sqrt{\varepsilon_{nm}+V_0-V_2+V_3+U_1} \qquad (16),$$

$$\delta_2(h) = \begin{vmatrix} \lambda_2(h) & s_2(h) \\ \lambda_3(h) & s_3(h) \end{vmatrix} = 0 \Leftrightarrow H_2 = -2 - J \pm \sqrt{\varepsilon_{nm}+V_0-V_2+V_3+U_1}$$

...etc.

By finding the *nth* term of the arithmetic progression in Eq. (16), we obtain;

$$H_n = -n - J \pm \sqrt{\varepsilon_{nm} + V_0 - V_2 + V_3 + U_1} \qquad (17).$$

By carrying out some algebraic manipulations on Eq. (17) and substituting the relevant expressions from (7) and (9), we get;

$$\varepsilon_{nm} = V_4 - V_5 - \eta_{nm} + \frac{1}{4}\left[\frac{V_0 - V_2 + V_3 + V_4 + V_5 - \eta_{nm} - \left(n + \frac{1}{2} + \sqrt{\frac{1}{4} + V_0 + V_1 + V_5 + U_0 + U_1 + \eta_m}\right)^2}{\left(n + \frac{1}{2} + \sqrt{\frac{1}{4} + V_0 + V_1 + V_5 + U_0 + U_1 + \eta_m}\right)}\right]^2 \qquad (18).$$

The energy is now obtained as follows;

$$E_{nm} = K + \frac{\hbar^2 (\alpha + \delta)^2}{\mu}(\Upsilon_0) - \frac{\hbar^2 (\alpha + \delta)^2}{4\mu}\left[\frac{\Upsilon_1 - (n + \tilde{J})^2}{(n + \tilde{J})}\right]^2 \qquad (19),$$

where

$$\tilde{J} = \frac{1}{2} + \sqrt{\frac{1}{4} + \frac{2\mu A}{\hbar^2} + \frac{2\mu B}{\hbar^2} + \frac{2\mu G}{\hbar^2(\alpha + \delta)} + \left(\frac{e\vec{B}m}{\hbar(\alpha + \delta)c} + \frac{e\vec{B}\phi_{AB}}{\hbar(\alpha + \delta)c\phi_0}\right) + \left(\frac{e\vec{B}}{2\hbar(\alpha + \delta)c}\right)^2 + \eta_m}$$

$$\Upsilon_0 = \eta_{nm} - \frac{\mu D/2(1+\tau)}{\hbar^2(\alpha + \delta)} + \frac{2\mu G}{\hbar^2(\alpha + \delta)} \qquad (20).$$

$$\Upsilon_1 = \frac{2\mu A}{\hbar^2} - \frac{\mu C}{\hbar^2(\alpha + \delta)} + \frac{\mu D/2}{\hbar^2(\alpha + \delta)} + \frac{\mu D/2(1+\tau)}{\hbar^2(\alpha + \delta)} + \frac{2\mu G}{\hbar^2(\alpha + \delta)} + \left(\frac{e\vec{B}}{2\hbar(\alpha + \delta)c}\right)^2 - \eta_{nm}$$

For completeness sake, we move to find the wave function of the system. In general, the differential equation we wish to solve should be transformed to a form given below, in accordance with the AIM formalism [36, 37]:

$$y''(x) = 2\left(\frac{ax^{N+1}}{1 - bx^{N+2}} - \frac{M+1}{x}\right)y'(x) - \frac{Wx^N}{1 - bx^{N+2}} y(x) \qquad (21),$$

where *a*, *b*, *M* and *W* are constants. The exact solutions for Eq. (21) is represented thus:

$$y(x) = (-1)^n C(N+2)^n (\sigma)_n {}_2F_1\left(-n, t+n; \sigma; bx^{N+2}\right) \qquad (22),$$

where,

$$(\sigma)_n = \frac{\Gamma(\sigma+n)}{\Gamma(\sigma)}, \sigma = \frac{2M+N+3}{N+2}, t = \frac{(2M+1)b+2a}{(N+2)b} \tag{23}.$$

By comparing Eq. (10) with Eq. (21), we can deduce that

$$M = J - \frac{1}{2}, t = 2(H+J), a = J\ \sigma = 2H+1,\ b=1, N=-1\ (\sigma_n) = \frac{\Gamma(2H+1+n)}{\Gamma(2H+1)} \tag{24}$$

and

$$f_{nm}(h) = (-1)^n C_2 \frac{\Gamma(2H+1+n)}{\Gamma(2H+1)} {}_2F_1(-n, 2(H+J)+n; 2H+1; h) \tag{25}.$$

It is therefore straightforward to deduce the corresponding unnormalized wave function as

$$R_{nm}(h) = (-1)^n C_2 \frac{\Gamma(2H+1+n)}{\Gamma(2H+1)} h^H (1-h)^J {}_2F_1(-n, 2(H+J)+n; 2H+1; h) \tag{26}.$$

3. **Thermo-magnetic properties of IUGE-HP**

The partition function of the IUGE-HP can be computed by a straightforward summation over all possible vibrational energy levels accessible to the system considered. With the help of Eq. (19), the partition function $Z(\beta)$ of the IUGE-HP at finite temperature $T$ is obtained as [38-41];

$$Z(\beta) = \sum_{n=0}^{n_{max}} e^{-\beta E_n} \tag{27},$$

with $\beta = \frac{1}{k_B T}$ and $k_B$ being the Boltzmann constant. To evaluate the partition function, the energy equation of Eq. (19) is rewritten as follows;

$$E_{nm} = P_0 - P_1 \left[\frac{\Upsilon_1 - (n+\tilde{J})^2}{(n+\tilde{J})}\right]^2 \tag{28}.$$

Here, we have defined the following parameters for mathematical simplicity:

$$P_0 = K + \frac{\hbar^2(\alpha+\delta)^2}{\mu}(\Upsilon_0),\ P_1 = \frac{\hbar^2(\alpha+\delta)^2}{4\mu}, \tag{29}.$$

Hence, by substituting Eq. (28) into Eq. (27), we obtain

$$Z(\beta) = \sum_{n=0}^{n_{max}} e^{-\beta\left(P_0 - P_1\left(\frac{\Upsilon_1 - (n+\tilde{J})^2}{(n+\tilde{J})}\right)^2\right)} \quad (30),$$

where

$$n_{max} = -\tilde{J} + \sqrt{P_0} \pm \sqrt{P_0 - \Upsilon_1} \quad (31),$$

represents the maximum quantum number. By converting the summation into integral in Eq. (30), we obtain

$$Z(\beta) = \int_0^{n_{max}} e^{-\beta\left(P_0 - P_1\left(\frac{\Upsilon_1 - (n+\tilde{J})^2}{(n+\tilde{J})}\right)^2\right)} dn \quad (32).$$

To evaluate the integral in Eq. (32), we carry out the process as follows;

$$Z(\beta) = \int_{\tilde{J}}^{n_{max}+\tilde{J}} e^{-\beta\left(P_1\rho^2 + \frac{R_0}{\rho^2} + R_1\right)} d\rho \quad (33).$$

where we have defined; $\rho = n + \tilde{J}$, $R_0 = Y_1^2 P_1$, $R_1 = -(2P_1\Upsilon_1 + P_0)$ and the limit of integral is given as: $\tilde{J} \leq \rho \leq n_{max} + \tilde{J}$.

The integral (33) is now evaluated using Mathematica software to obtain partition function, $Z(\beta)$ as follows;

$$Z(\beta) = -\frac{e^{R_1\beta - 2\sqrt{-P_1\beta}\sqrt{-R_0\beta}}\sqrt{\pi}}{4\sqrt{-P_1\beta}} \begin{pmatrix} Erf\left[\tilde{J}\sqrt{-P_1\beta} - \frac{\sqrt{-R_0\beta}}{\tilde{J}}\right] + e^{4\sqrt{-P_1\beta}\sqrt{-R_0\beta}} Erf\left[\tilde{J}\sqrt{-P_1\beta} - \frac{\sqrt{-R_0\beta}}{\tilde{J}}\right] \\ -Erf\left[(\tilde{J}+n_{max})\sqrt{-P_1\beta} - \frac{\sqrt{-R_0\beta}}{(\tilde{J}+n_{max})}\right] - \\ e^{4\sqrt{-P_1\beta}\sqrt{-R_0\beta}} Erf\left[(\tilde{J}+n_{max})\sqrt{-P_1\beta} - \frac{\sqrt{-R_0\beta}}{(\tilde{J}+n_{max})}\right] \end{pmatrix} \quad (34).$$

It is well-known that in order to investigate the thermodynamics properties of a given system, then it is necessary to calculate its partition function as given in equation (34). Equation (34) in statistical physics is a function of temperature and plays a role of a distribution and once equation (34) is known, then other thermodynamics properties can be obtained from it such as entropy, internal energy, specific heat capacity and Helmholtz free energy. Here, *Erf* represents the error function [38-41]. By employing Eq. (34), different thermodynamic and magnetic properties of the IUGE-HP such as the free energy, the entropy, total energy, the specific heat, magnetization and magnetic susceptibility can be obtained for diatomic molecules system using the following expressions [38-41];

$$F(\beta) = -\frac{1}{\beta}\ln Z(\beta); \quad U(\beta) = -\frac{d\ln Z(\beta)}{d\beta};$$

$$S(\beta) = \ln Z(\beta) - \beta\frac{d\ln Z(\beta)}{d\beta}; \quad C(\beta) = \beta^2\frac{d^2\ln Z(\beta)}{d\beta^2}; \quad (35).$$

$$M(\beta) = \frac{1}{\beta}\left(\frac{1}{Z(\beta)}\right)\left(\frac{\partial}{\partial\vec{B}}Z(\beta)\right); \quad \chi_m(\beta) = \frac{\partial M(\beta)}{\partial\vec{B}}$$

4. **Discussion and Application of Results**

A. **Energy spectra**

Tables (1-3) shows energy values of the IUGE-HP under the influence of magnetic and AB fields with selected values of magnetic quantum numbers for $\tau = 0, -1$ and $1$. We observe that when both the magnetic and AB fields are absent $(\vec{B} = \Phi_{AB} = 0)$, there is an existence some quasi - degeneracy in the energy levels of the considered system. By exposing the system to the solitary effect of the magnetic field $(\vec{B} \neq 0, \Phi_{AB} = 0)$, the energy results is reduced and becomes attractive and also degeneracy is removed. When we apply the AB field alone $(\vec{B} = 0, \Phi_{AB} \neq 0)$, quasi-degeneracy is present and the system is less bounded as we notice that the system approaches the continuum state as the values are positive. The combined effect of both fields is strong and therefore, there is a major downward shift in the energy levels of the system. The combined effect shows that the system is highly attractive and bounded. On the other hand, the control parameters considered tend to create a downward shift, making the energy levels of the system to decrease as $\tau$ increases.

B. **Partition function**

The plots of the partition function of IUGE-HP with the reduced temperature $\beta$, external magnetic, AB fields and control parameter $\tau$ are displayed in Fig.1 . The variation of the partition versus $\beta$ with varying $n_{max}$ is shown in Fig.1(a). The plot of the partition versus $\vec{B}$ for varying $\beta$ is shown in Fig.1(b). The plot of the partition function versus $\phi_{AB}$ with varying $\beta$ is shown in Fig.1 (c). The variation of the partition versus the control parameter $\tau > 0$ varying $\beta$ is displayed in Fig.1(d) and the plot of the partition function versus the control parameter $\tau < 0$ varying $\beta$ are shown in Fig.1€. In Fig. 1(a), the partition function increases with increasing $\beta$. Fig. 1(b) shows that the partition function rises with increasing magnetic field. Fig. 1(c), the partition function reduces with increasing AB field. In Fig. 1 (d), the partition function increases with increasing values of the control parameter for $\tau > 0$. In Fig. 1 (e), the partition function increases with increasing values of the control parameter for $\tau < 0$

### C. Free Energy

The plots of the free energy of the system of the IUGE-HP with reduced temperature, magnetic AB fields and the control parameter are shown in Fig.2. We displayed in Fig.2(a) the free energy versus $\beta$ with varying $n_{max}$. The variation free energy versus $\vec{B}$ for varying $\beta$ is shown in Fig.2.(b). The plot of the free energy versus $\phi_{AB}$ with varying $\beta$ is displayed in Fig.2.(c). Also, the plots of the free energy versus the control parameter $\tau > 0$ varying $\beta$ is shown in Fig.2(d) and the plot of the free energy versus the control parameter $\tau < 0$ varying $\beta$ is displayed in Fig.2€. In Fig. 2(a), the free energy increases with increased $\beta$. Fig. 2(b) shows that the free energy decreases with increased magnetic field. Fig. 2(c), the free energy increases with increased AB field. In Fig. 2 (d), the free energy decreases with increased values of the control parameter for $\tau > 0$. In Fig. 2 (e), the free energy decreases with increased values of the control parameter for $\tau < 0$.

### D. Entropy

The variation of the entropy of the IUGE-HP with temperature, external magnetic field, AB fields and control parameter are shown in Fig.3. The behavior of the entropy versus $\beta$ with varying $n_{max}$ is displayed in Fig.3(a) The entropy versus $\vec{B}$ for varying $\beta$ plot is shown in Fig.3(b). The variation of the entropy versus $\phi_{AB}$ for varying $\beta$ is shown in Fig.3(c). Similarly, the plots of the entropy versus the control parameter $\tau > 0$ for varying $\beta$ is shown in Fig. 3(d) and entropy versus the control parameter $\tau < 0$ for varying $\beta$ is shown in Fig.3(e) In Fig. 3(a), the entropy decreases with increasing $\beta$. Fig. 3(b) shows that the entropy decreases with increasing magnetic field. In Fig. 3(c), the entropy increases with increasing AB field. In Fig. 3 (d), the entropy decreases with increasing values of the control parameter for $\tau > 0$. In Fig. 3 (e), the entropy decreases with increasing values of the control parameter for $\tau < 0$.

### E. Mean Energy

The plots of the mean energy of IUGE-HP with temperature, external magnetic , AB fields and control parameter are shown in Figs.4(a-e). The plot of mean energy versus $\beta$ for varying $n_{max}$ is displayed in Fig.4(a). The behaviour of the mean energy versus $\vec{B}$ for varying $\beta$ is shown in Fig.4(b). The variation of the mean energy versus $\phi_{AB}$ for varying $\beta$ is shown in Fig.4(c). The plots of the mean energy versus the control parameter $\tau > 0$ varying $\beta$ is shown in Fig.4(d) and mean energy versus the control parameter $\tau < 0$ varying $\beta$ is displayed in Fig.4(d). In Fig. 4(a), the average energy decreases with increasing $\beta$. Fig. 4(b) shows that the average energy decreases with increasing magnetic field. In Fig. 4(c), the average energy increases with increasing AB field.

In Fig. 4 (d), the average energy decreases with increasing values of the control parameter for $\tau > 0$. In Fig. 4 (e), the average energy decreases with increasing values of the control parameter for $\tau < 0$.

### F. Specific Heat Capacity

Fig. 5 shows the plots of Specific heat capacity of IUGE-HP with magnetic and AB fields ; (a) versus $\beta$ varying $n_{max}$, (b) versus $\vec{B}$ for varying $\beta$, (c) versus $\phi_{AB}$ varying $\beta$, (d) versus the control parameter $\tau > 0$ varying $\beta$, (e) versus the control parameter $\tau < 0$ varying $\beta$. In Fig. 5(a), the Specific heat capacity increases with increasing $\beta$. Fig. 5(b) shows that the Specific heat capacity increases with increasing magnetic field. In Fig. 5(c), the Specific heat capacity decreases with increasing AB field. Fig. 5 (d) shows that the Specific heat capacity increases with increasing values of the control parameter for $\tau > 0$. In Fig. 5 (e), shows that the Specific heat capacity increases with increasing values of the control parameter for $\tau < 0$. Fig. 6 shows the plots of magnetization of IUGE-HP with magnetic and AB fields; (a) versus $\beta$ varying $n_{max}$, (b) versus $\vec{B}$ for varying $\beta$, (c) versus $\phi_{AB}$ varying $\beta$, (d) versus the control parameter $\tau > 0$ varying $\beta$, (e) versus the control parameter $\tau < 0$ varying $\beta$.

### G. Magnetization

In Fig. 6(a), we plot the variation of the magnetization with reduced temperature $\beta$. Fig.6 (a) shows that the magnetization M increases with increased in temperature $\beta$. Also, we plotted the variation of magnetization with the external magnetic field B in Fig.6(b). As shown in Fig. 6(b), it is observed that the magnetization increases with increased in external magnetic field. The variation of the magnetization with the AB field is displayed in Fig. 6 (c). Here, the magnetization decreases with increased in AB field. The effects of the control parameter on the magnetization are shown in Figs. 6(d&e) for $\tau > 0$ and $\tau < 0$ respectively. In Fig. 6 (d), the magnetization increases with increased in the values of the control parameter for $\tau > 0$ and also in Fig. 6 (e), the magnetization increases with increased in the values of the control parameter for $\tau < 0$.

### H. Magnetic Susceptibility

In Fig. 7 we plotted the variation of the magnetic susceptibility of IUGE-HP with temperature, magnetic fields AB fields and control parameters. In Fig.7 (a) the variation of magnetic susceptibility versus $\beta$ for varying $n_{max}$ are displayed. Also, in Fig.7 (b) the plot of the magnetic susceptibility versus $\vec{B}$ for varying $\beta$ are shown. The plot of the magnetic susceptibility versus $\phi_{AB}$ for varying $\beta$ is shown in Fig.7(c). The plots of the magnetic susceptibility versus the control parameter $\tau$ are displayed in Fig.7 (d) versus the control parameter $\tau > 0$ varying $\beta$ and Fig.7 (e) versus the control parameter $\tau < 0$ varying $\beta$. In Fig. 7(a), the magnetic susceptibility increases with increased in $\beta$. Fig. 7(b) shows that the magnetic susceptibility increases with increased magnetic field. In Fig. 7 (c), the magnetic susceptibility decreases with increased AB field. Fig. 7 (d) shows that the magnetic susceptibility increases with increased in values of the control parameter for $\tau > 0$. In Fig. 7 (e), the magnetic susceptibility increases with increased in the values of the control parameter for $\tau < 0$.

## 5. Conclusions

In this article, we solve the SE in the presence of magnetic and AB fields for the IUGE-HP using the AIM formalism. We obtain the eigenvalue equation and eigen-function for the IUGE-HP. We calculated the energy levels of the quantum state in the present of external magnetic as well as the AB flux field. It is well-known that the AB effect is a quantum mechanical phenomenon where a charged particle is affected by an electromagnetic field despite being confined to a region in which both the magnetic field and electric field are zero[42-43]. We investigated the influence of the external magnetic fields, AB fields and the control parameter on the energy spectra and thermal properties of the IUGE-HP. As shown in Tables 1-3, It was found out that the $\vec{B}$ and AB fields remove degeneracy for various potential parameters of IUGE-HP. However, the present of the AB field was observed to remove the degeneracy more better than the present of the external magnetic field only. In addition, we investigated the magnetization, magnetic susceptibility and thermal properties of the system These thermodynamic properties of IUGE-HP potential were thoroughly investigated in the presence of both external magnetic and AB fields. . Finally, our research findings could also be applied in condensed matter physics, atomic and chemical physics for position dependent mass systems[44-47].

**Acknowledgments**

The authors NAA and SKE acknowledged the financial support from Taif University Researchers Supporting Project number (TURSP-2020/247), Taif University, Taif, Saudi Arabia.

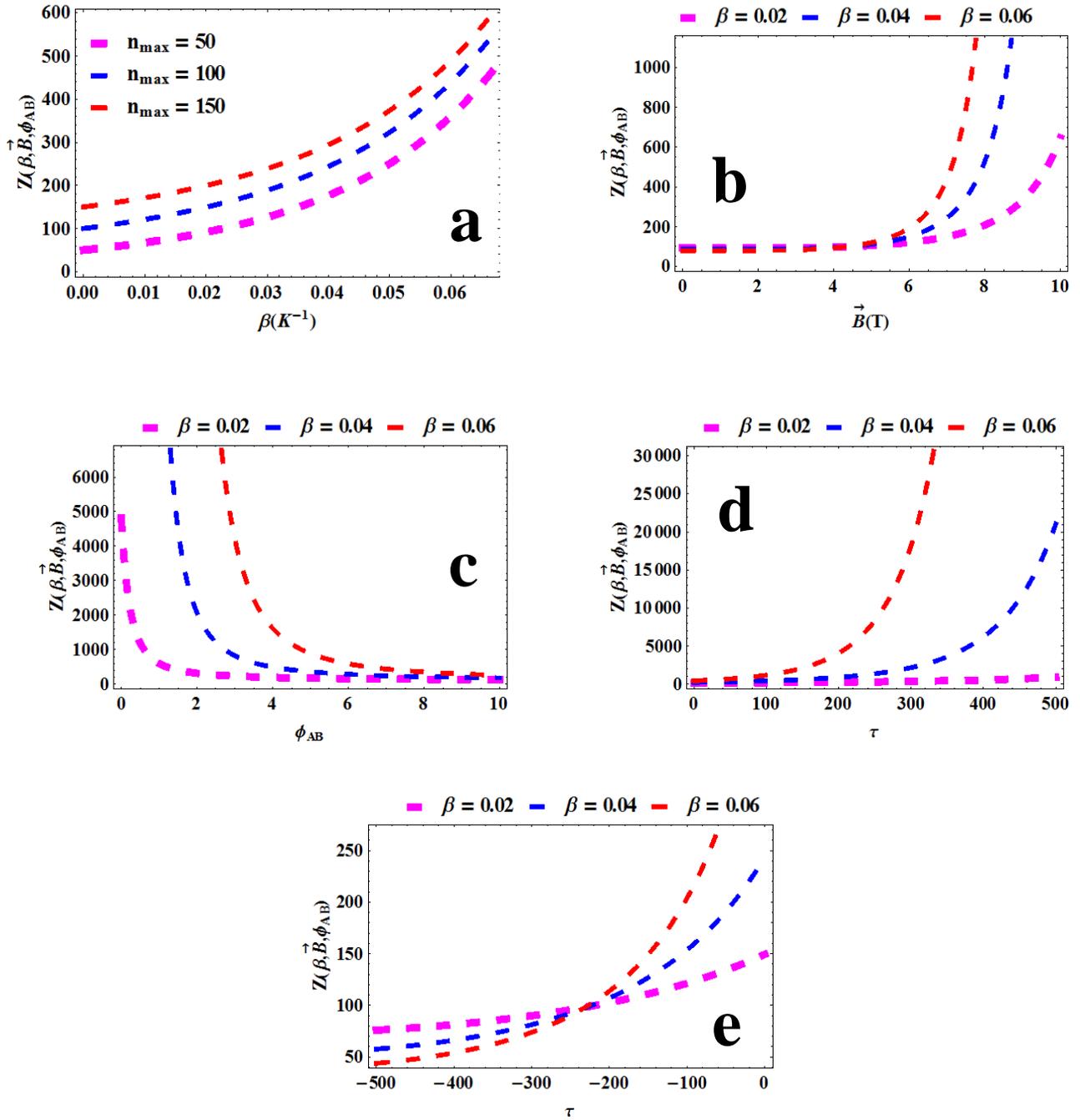

**Figure 1**: Plots of partition function of IUGE-HP with magnetic and AB fields; (a) versus $\beta$ varying $n_{max}$, (b) versus $\vec{B}$ for varying $\beta$, (c) versus $\phi_{AB}$ varying $\beta$, (d) versus the control parameter $\tau > 0$ varying $\beta$, (e) versus the control parameter $\tau < 0$ varying $\beta$.

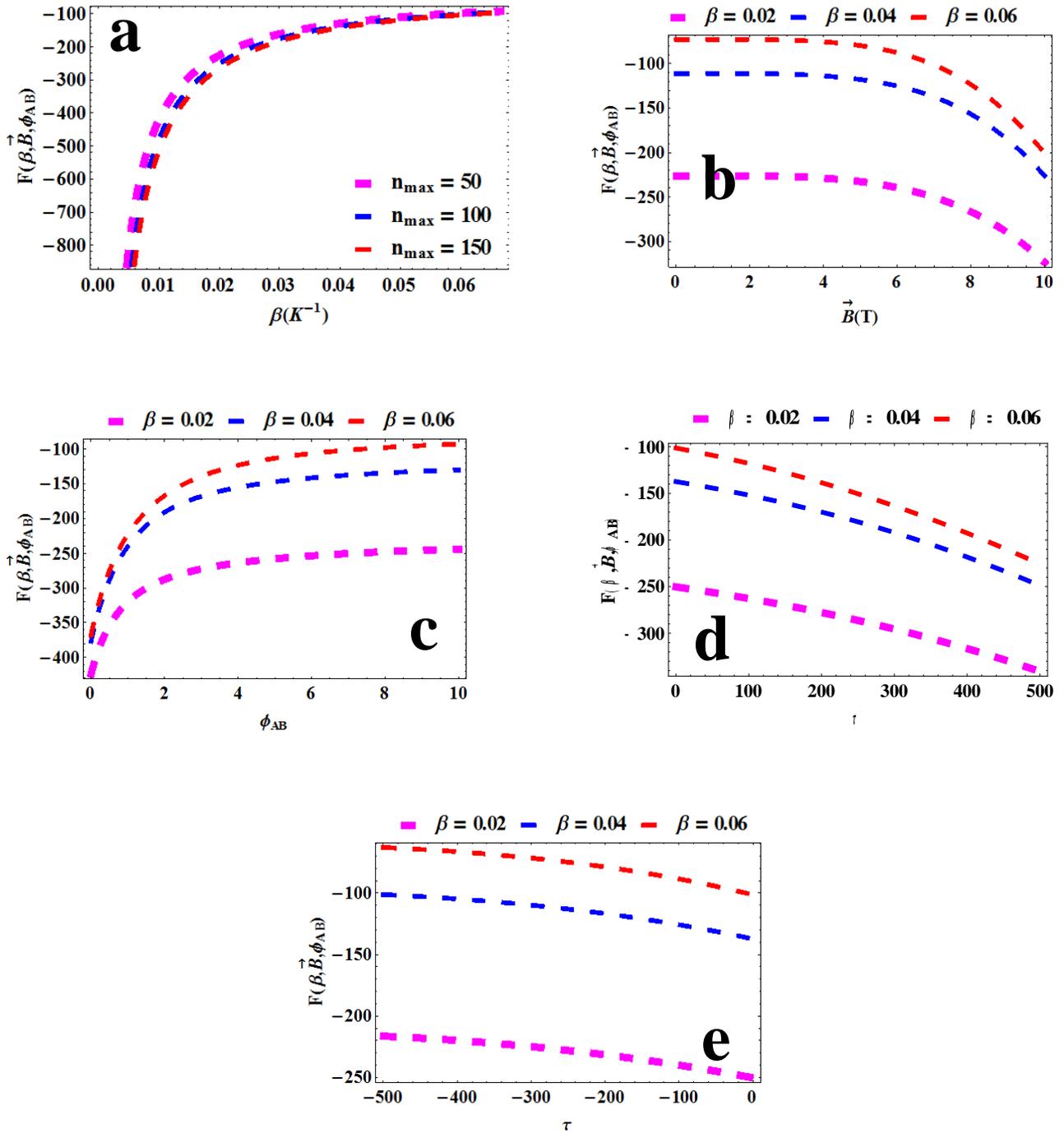

**Figure 2**: Plots of Free energy of IUGE-HP with magnetic and AB fields; (a) versus $\beta$ varying $n_{max}$, (b) versus $\vec{B}$ for varying $\beta$, (c) versus $\phi_{AB}$ varying $\beta$, (d) versus the control parameter $\tau > 0$ varying $\beta$, (e) versus the control parameter $\tau < 0$ varying $\beta$.

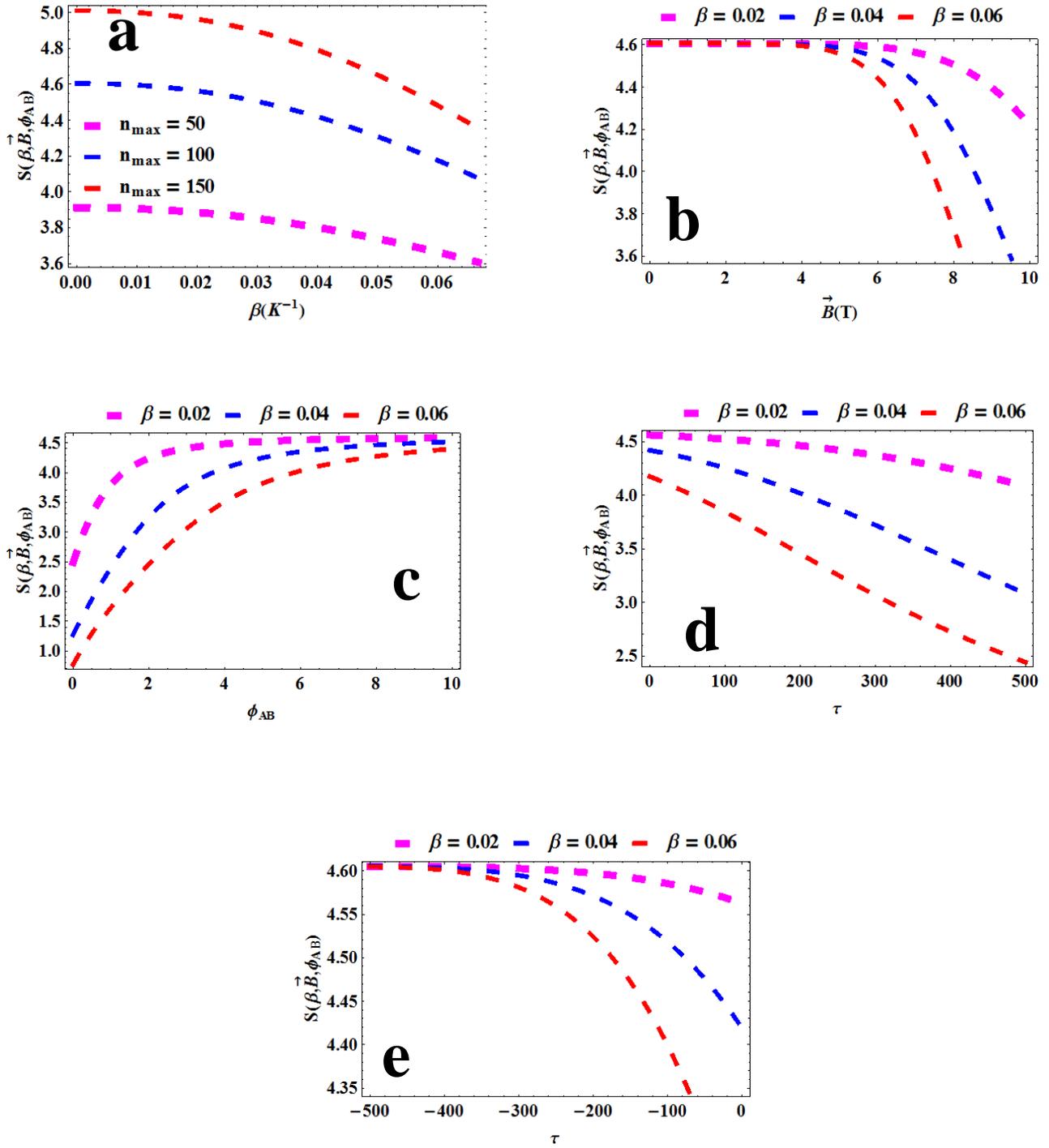

**Figure 3**: Plots of Entropy of IUGE-HP with magnetic and AB fields; (a) versus $\beta$ varying $n_{max}$, (b) versus $\vec{B}$ for varying $\beta$, (c) versus $\phi_{AB}$ varying $\beta$, (d) versus the control parameter $\tau > 0$ varying $\beta$, (e) versus the control parameter $\tau < 0$ varying $\beta$.

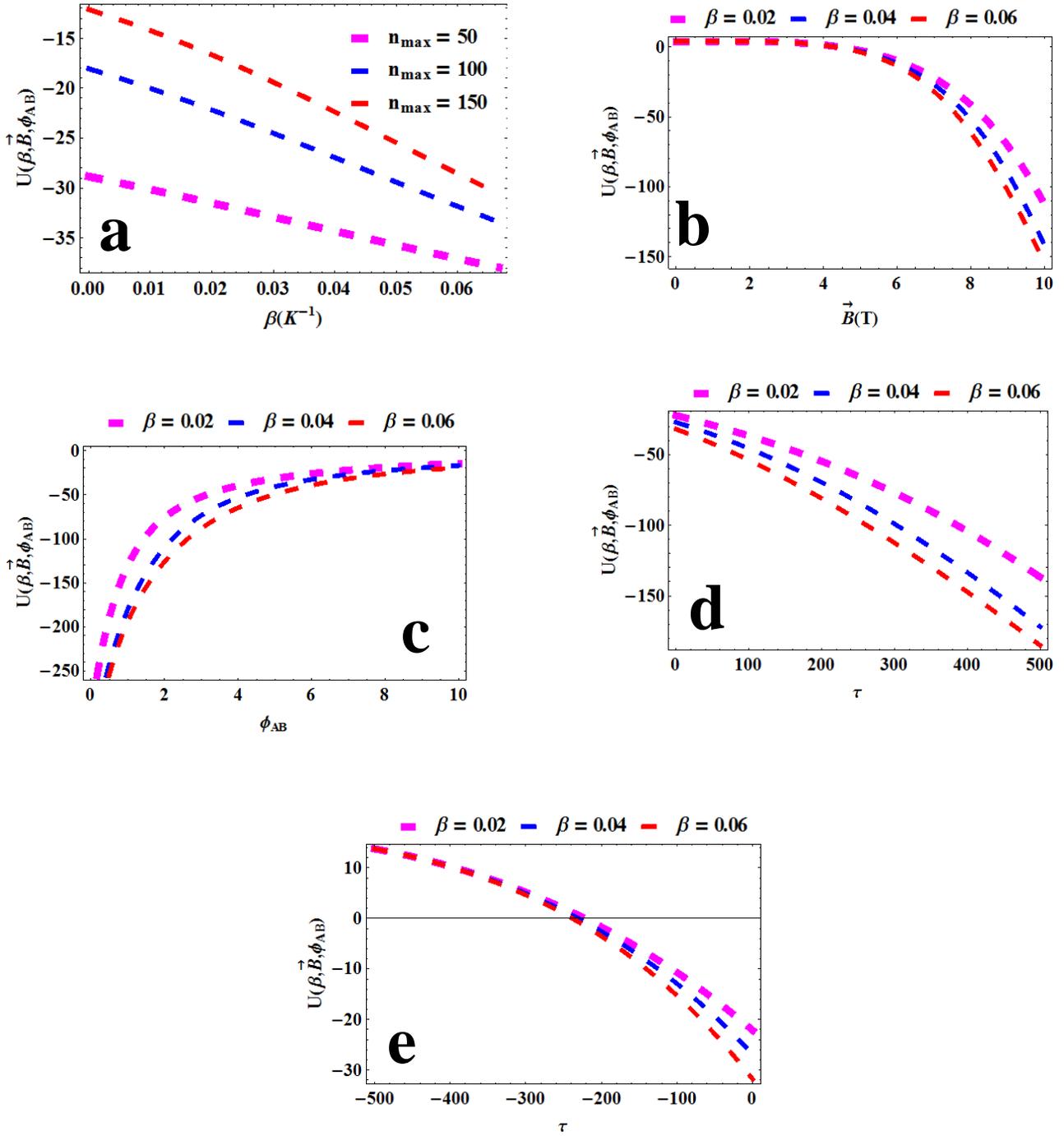

**Figure 4**: Plots of Average or mean energy of IUGE-HP in magnetic and AB fields; (a) versus $\beta$ varying $n_{max}$, (b) versus $\vec{B}$ for varying $\beta$, (c) versus $\phi_{AB}$ varying $\beta$, (d) versus the control parameter $\tau > 0$ varying $\beta$, (e) versus the control parameter $\tau < 0$ varying $\beta$.

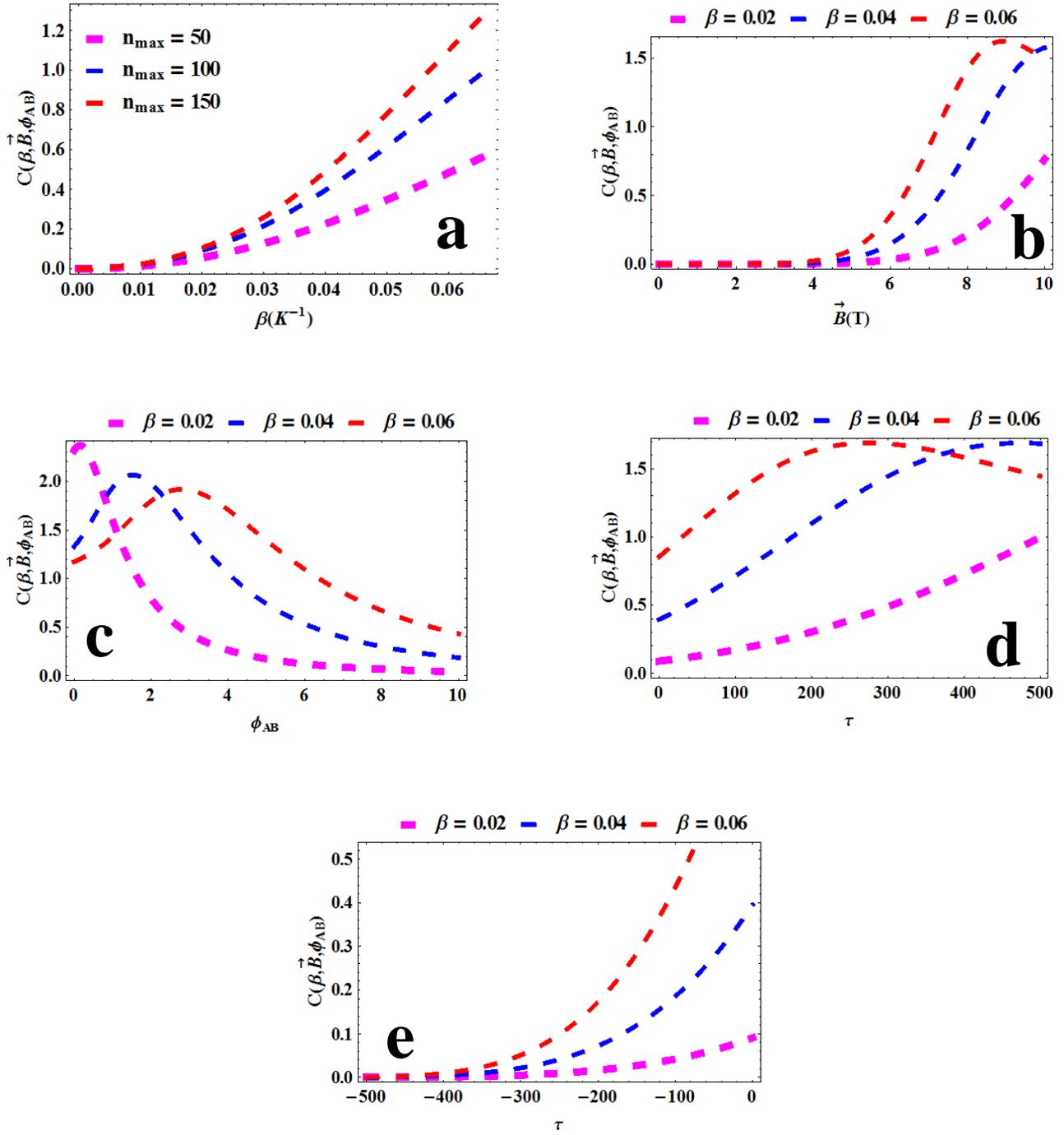

**Figure 5**: Plots of Specific heat capacity of IUGE-HP with magnetic and AB fields; (a) versus $\beta$ varying $n_{max}$, (b) versus $\vec{B}$ for varying $\beta$, (c) versus $\phi_{AB}$ varying $\beta$, (d) versus the control parameter $\tau > 0$ varying $\beta$, (e) versus the control parameter $\tau < 0$ varying $\beta$.

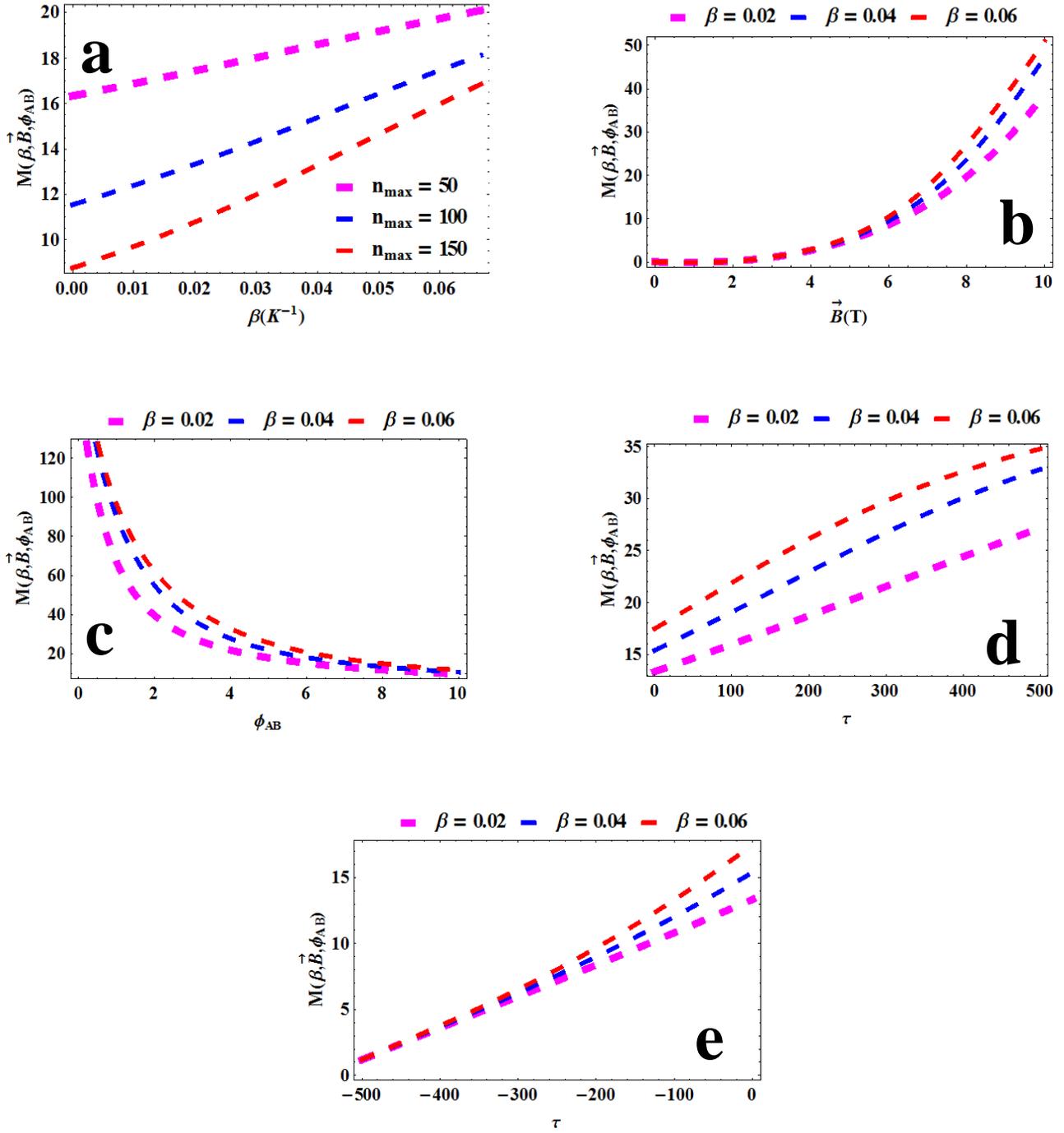

**Figure 6**: Plots of magnetization of IUGE-HP in magnetic and AB fields; (a) versus $\beta$ varying $n_{max}$, (b) versus $\vec{B}$ for varying $\beta$, (c) versus $\phi_{AB}$ varying $\beta$, (d) versus the control parameter $\tau > 0$ varying $\beta$, (e) versus the control parameter $\tau < 0$ varying $\beta$.

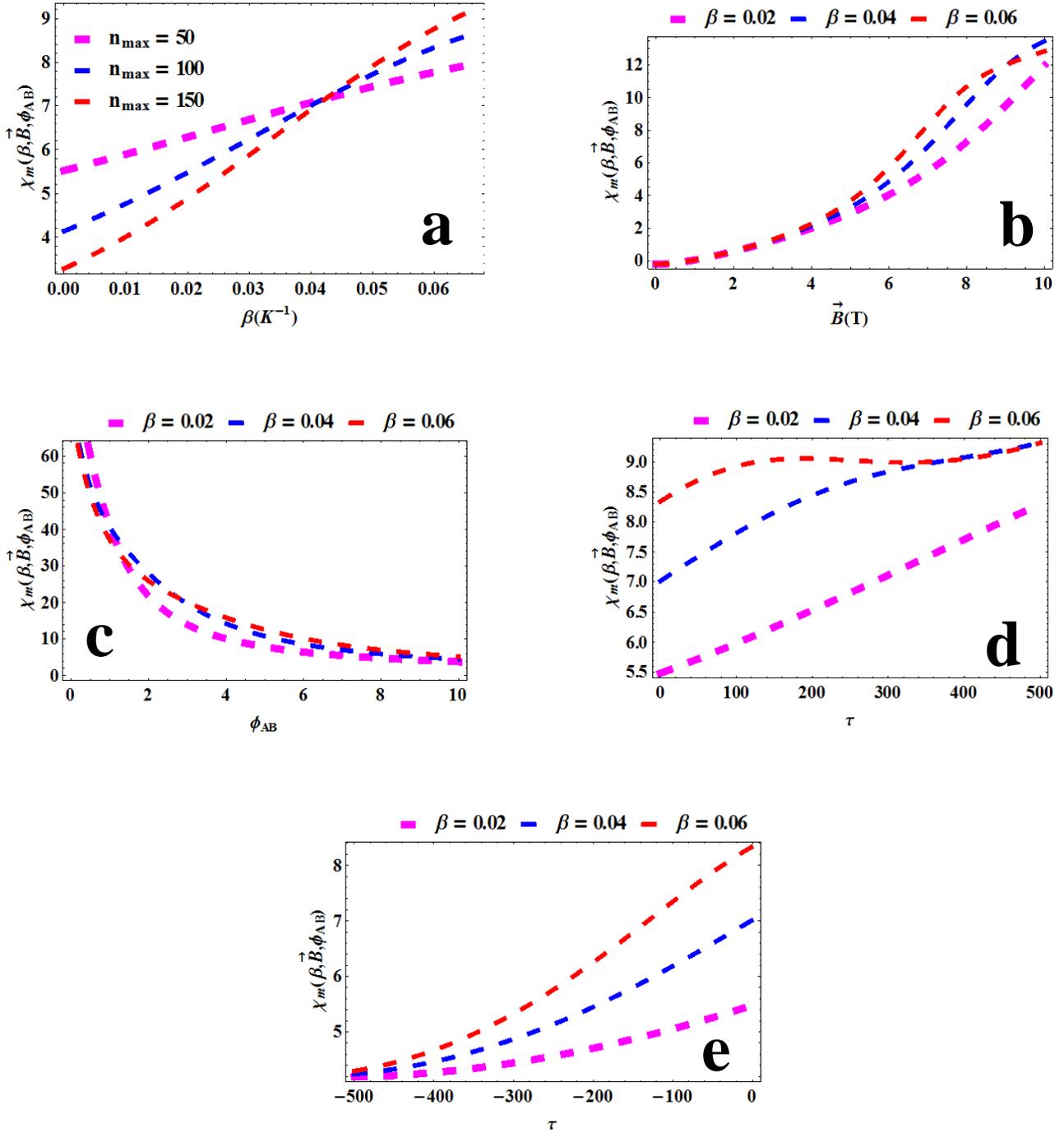

**Figure 7**: Plots of magnetic susceptibility of IUGE-HP with magnetic and AB fields; (a) versus $\beta$ varying $n_{max}$, (b) versus $\vec{B}$ for varying $\beta$, (c) versus $\phi_{AB}$ varying $\beta$, (d) versus the control parameter $\tau > 0$ varying $\beta$, (e) versus the control parameter $\tau < 0$ varying $\beta$.

**Table 1**: Energy values for the IUGE-HP under the combined influence of magnetic and AB fields with various values of magnetic and principal quantum numbers for $\tau = -1$.

| $m$ | $n$ | $\vec{B}=0, \phi_{AB}=0$ | $\vec{B}=7, \phi_{AB}=0$ | $\vec{B}=0, \phi_{AB}=7$ | $\vec{B}=7, \phi_{AB}=7$ |
|---|---|---|---|---|---|
| 0 | 0 | 4.05994 | -372.02700 | 4.06060 | -53.66680 |
|   | 1 | 4.05973 | -348.85800 | 4.06034 | -52.04890 |
|   | 2 | 4.05935 | -327.73400 | 4.05992 | -50.49220 |
|   | 3 | 4.05883 | -308.42100 | 4.05936 | -48.99370 |
| -1 | 0 | 4.05984 | -1365.79000 | 4.06050 | -62.23700 |
|   | 1 | 4.05964 | -1213.82000 | 4.06025 | -60.28280 |
|   | 2 | 4.05927 | -1085.70000 | 4.05984 | -58.40710 |
|   | 3 | 4.05875 | -976.68800 | 4.05928 | -56.60570 |
| 1 | 0 | 4.06003 | -212.42400 | 4.06069 | -46.92170 |
|   | 1 | 4.05982 | -202.01500 | 4.06043 | -45.55380 |
|   | 2 | 4.05944 | -192.31900 | 4.06000 | -44.23490 |
|   | 3 | 4.05891 | -183.27400 | 4.05943 | -42.96270 |

**Table 2**: Energy values for the IUGE-HP under the combined influence of magnetic and AB fields with various values of magnetic and principal quantum numbers for $\tau = 0$.

| $m$ | $n$ | $\vec{B}=0, \phi_{AB}=0$ | $\vec{B}=7, \phi_{AB}=0$ | $\vec{B}=0, \phi_{AB}=7$ | $\vec{B}=7, \phi_{AB}=7$ |
|---|---|---|---|---|---|
| 0 | 0 | 4.03885 | -373.28000 | 4.03974 | -53.88400 |
|   | 1 | 4.03947 | -350.03500 | 4.04029 | -52.26090 |
|   | 2 | 4.03983 | -328.84300 | 4.04059 | -50.69910 |
|   | 3 | 4.03997 | -309.46700 | 4.04067 | -49.19570 |
| -1 | 0 | 4.03872 | -1370.28000 | 4.03961 | -62.48220 |
|   | 1 | 4.03935 | -1217.81000 | 4.04017 | -60.52170 |
|   | 2 | 4.03972 | -1089.27000 | 4.04048 | -58.63980 |
|   | 3 | 4.03987 | -979.90700 | 4.04057 | -56.83260 |
| 1 | 0 | 4.03897 | -213.15800 | 4.03986 | -47.11700 |
|   | 1 | 4.03959 | -202.71500 | 4.04040 | -45.74460 |
|   | 2 | 4.03994 | -192.98800 | 4.04070 | -44.42140 |
|   | 3 | 4.04007 | -183.91300 | 4.04077 | -43.14500 |

**Table 3**: Energy values for the IUGE-HP under the combined influence of magnetic and AB fields with various values of magnetic and principal quantum numbers for $\tau = 1$.

| $m$ | $n$ | $\vec{B}=0, \phi_{AB}=0$ | $\vec{B}=7, \phi_{AB}=0$ | $\vec{B}=0, \phi_{AB}=7$ | $\vec{B}=7, \phi_{AB}=7$ |
|---|---|---|---|---|---|
| 0  | 0 | 4.01461 | -374.53400  | 4.01576 | -54.10160 |
|    | 1 | 4.01629 | -351.21500  | 4.01736 | -52.47320 |
|    | 2 | 4.01761 | -329.95300  | 4.01859 | -50.90640 |
|    | 3 | 4.01860 | -310.51500  | 4.01950 | -49.39810 |
| -1 | 0 | 4.01444 | -1374.77000 | 4.01560 | -62.72780 |
|    | 1 | 4.01614 | -1221.80000 | 4.01721 | -60.76090 |
|    | 2 | 4.01747 | -1092.85000 | 4.01845 | -58.87290 |
|    | 3 | 4.01847 | -983.13100  | 4.01937 | -57.05980 |
| 1  | 0 | 4.01477 | -213.89300  | 4.01593 | -47.31250 |
|    | 1 | 4.01645 | -203.41600  | 4.01751 | -45.93570 |
|    | 2 | 4.01775 | -193.65700  | 4.01873 | -44.60820 |
|    | 3 | 4.01873 | -184.55300  | 4.01963 | -43.32770 |

# REFERENCES

[1] W. Greiner, Relativistic Quantum Mechanics: Wave Equations (Springer, Berlin, 2000).

[2] L.D. Landau and E.M. Lifshitz, Quantum Mechanics, Non- Relativistic Theory (Pergamon, New York, 1977).

[3] L.I. Schiff, Quantum Mechanics (McGraw Hill, New York, 1995).

[4] P.A. Dirac, The Principles of Quantum Mechanics (Clarendon Press, Oxford, 1958).

[5] P.O. Okoi, C.O. Edet and T.O.Magu, Rev. Mex. Fis. **66** (2020) 1.

[6] C.O. Edet, U.S. Okorie, A.T. Ngiangia and A.N. Ikot, Ind. J. Phys. **94** (2019) 425.

[7] C.O. Edet and P.O. Okoi, Rev. Mex. Fis. **65** (2019) 333

[8] U.S. Okorie, A.N. Ikot, C.O. Edet, I.O. Akpan, R. Sever and R. Rampho, J. Phys. Commun. **3** (2019) 095015.

[9] C.O. Edet, U.S. Okorie, G. Osobonye, A.N. Ikot, G.J. Rampho and R. Sever, J. Math. Chem. **58** (2020) 989.

[10] C.O. Edet, P.O. Okoi, A.S. Yusuf, P.O. Ushie and P.O. Amadi, Ind. J. Phys. **95** (2021) 471

[11] C. O. Edet, K. O. Okorie, H. Louis and N. A. Nzeata- Ibe, Indian J Phys **94** (2020) 243

[12] C.O. Edet, P.O. Okoi and S.O. Chima, Rev. Bras. Ens. Fís. **42** (2020) e20190083.

[13] A.N. Ikot, U.S. Okorie, A.T. Ngiangia, C.A. Onate, C.O. Edet, I.O. Akpan and P.O. Amadi, Eclét Quím J. **45** (2020) 65.

[14] R. H. Parmar, Few-Body Syst. **61** (2020) 39.

[15] A.N. Ikot, C.O. Edet, P.O. Amadi, U.S. Okorie, G.J. Rampho, H.Y. Abdullah, Eur. Phys. J. D **74** (2020) 159

[16] A.N. Ikot, U.S. Okorie, G. Osobonye, P.O. Amadi, C.O. Edet, M.J. Sithole, G.J. Rampho, R. Sever, Heliyon 6 (2020) e03738.

[17] C. O. Edet, P. O. Amadi, M. C. Onyeaju, U. S. Okorie, R. Sever, G. J. Rampho, Hewa Y. Abdullah, Idris H. Salih and A. N. Ikot, J. Low Temp. Phys. **202** (2021) 83

[18] G. J. Rampho, A. N. Ikot, C. O. Edet & U. S. Okorie, Molecular Physics, **119** (2021) 5, DOI: 10.1080/00268976.2020.1821922

[19] P. Ghosh and D. Nath, Int. J. Quant. Chem., 1–19 (2020).https://doi.org/10.1002/qua.26153

[20] B.T. Mbadjoun, J.M. Ema, P.E. Abiama, G.H. Ben-Bolie and P. Owono Ateba, Mod. Phys Lett. A. **1** (2020) 2050092

[21] H. Karayer, Eur. Phys. J. Plus. **135** (2020) 70.

[22] N. Ferkous and A. Bounames, Commun. Theor. Phys. **59** (2013) 679.

[23] S.M. Ikhdair and B.J. Falaye, J. Ass. Arab Univ. Basic Appl. Sci. **16** (2014) 1.

[24] M. Eshghi and H. Mehraban, Eur. Phys. J. Plus. **132** (2017) 121

[25] M. Eshghi, H.Mehraban and S.M. Ikhdair, Eur. Phys. J. A. **52** (2016) 201.

[26] R. Khordad, M.A. Sadeghzadeh and A.M. Jahan-abad, Commun. Theor. Phys. **59** (2013) 655.

[27] P. Koscik and A. Okopinska, J. Phys.Math. Theor. **40** (2007) 1045.

[28] M. Aygun, O. Bayrak, I. Boztosun andY. Sahin, Eur. Phys. J. D. 66, 35 (2012).

[29] M.K. Elsaid, A. Shaer, E. Hjaz and M.H. Yahya, Chin. J. Phys. (2020). https://doi.org/10.1016/j.cjph.2020.01.002

[30] S. Gumber, M. Kumar, M. Gambhir, M. Mohan and P.K. Jha, Can. J. Phys. **93** (2015) 1.

[31] R. Khordad and H.R.R. Sedehi, J. Low Temp. Phys. **190** (2017) 200.

[32] R. Khordad, Int. J. Thermophys. **34** (2013) 1148.

[33] R. Khordad, H. Bahramiyan and H.R.R. Sedehi, Opt. Quant. Elect. **50** (2018) 294.

[34] M. Eshghi, R. Sever and S.M. Ikhdair, Chin. Phys. B. **27** (2018) 020301.

[35] R.L. Greene and C. Aldrich, Phys. Rev. A. **143** (1976) 2363

[36] H. Çiftçi, R L. Hall and N. Saad. Phys. Lett. A **340** (2005) 388

[37] B J Falaye Cent. Eur. J. Phys. **10** (2012) 960

[38] A. Boumali, J. Math. Chem. **56** (2018)1656.

[39] A. Bera, A. Ghosh and M. Ghosh, J. Magnet. Mag. Mater. **484** (2019) 391.

[40] C.S. Jia, L.H. Zhang and C.W. Wang, Chem. Phys. Lett. **667** (2017) 211.

[41] C.S. Jia, C.W. Wang, L.H. Zhang, X.L. Peng, R. Zeng and X.T. You, Chem. Phys. Lett. **676** (2017) 150

[42] B. J. Falaye, G. H. Sun, R. S. Ortigoza and S. H. Dong, Phys. Rev. E 93(2016)053201

[43] J. Bosse, Z.Naturforch,**72** (2017) 717

[44] R. A. El-Nabulsi, Few Body Syst. **61** (2020) 37

[45] M. Eshghi, R. Sever and S. M. Ikhdair, Chin. Phys. B **27** (2018) 020301

[46] S. H. Dong, J. J. Pena, C. Pacheco-Garcia and J. Garcia-Ravelo, Mod. Phys. Lett. A **22** (2007)1039

[47] R. A. El-Nabulsi, Euro. Phys. J. Plus **135** (2020) 693